\documentclass[journal,twoside,web]{ieeecolor}
\usepackage{tmi}
\usepackage{cite}
\usepackage{amsmath,amssymb,amsfonts}
\usepackage{mathtools}
\usepackage{algorithmic}
\usepackage{graphicx}
\usepackage{textcomp}
\usepackage{epsfig}
\usepackage{booktabs}
\usepackage{multirow}
\usepackage{hyperref}
\usepackage{url}
\usepackage{tabularx}

\usepackage{subfigure}
\usepackage[dvipsnames]{xcolor}
\hypersetup{
    colorlinks=true,
    linkcolor=blue,
    filecolor=magenta,
    urlcolor=cyan,
}
\usepackage{color}
\newcommand{\rednew}[1]{{\color{black}{#1}}}
\newcommand{\red}[1]{{\color{black}{#1}}}
\newcommand{\black}[1]{{\color{black}{#1}}}
\def\BibTeX{{\rm B\kern-.05em{\sc i\kern-.025em b}\kern-.08em
    T\kern-.1667em\lower.7ex\hbox{E}\kern-.125emX}}
\begin{document}
\title{Graph Convolution Based Cross-Network Multi-Scale Feature Fusion for Deep Vessel Segmentation}
\author{Gangming Zhao,
        Kongming Liang,
        Chengwei Pan,
        Fandong Zhang,
        Xianpeng Wu,\\
        Xinyang Hu,
        and Yizhou Yu, \IEEEmembership{Fellow, IEEE}
\thanks{This work was funded in part by National Key Research and Development Program of China (No. 2019YFC0118101), National Natural Science Foundation of China (Grant Nos. 62141605 and 82072005), Key Program of Beijing Municipal Natural Science Foundation (No.7191003), and Zhejiang Province Key Research \& Development Program (No. 2020C03073). \textit{(Corresponding authors: Yizhou Yu and Xinyang Hu.)}}
\thanks{Gangming Zhao and Yizhou Yu are with the Department of Computer Science, The University of Hong Kong, Hong Kong (e-mail: gmzhao@connect.hku.hk, yizhouy@acm.org).}
\thanks{Kongming Liang is with Pattern Recognition and Intelligent System Laboratory, School of Artificial Intelligence, Beijing University of Posts and Telecommunications, Beijing, China (e-mail: liangkongming@bupt.edu.cn).}
\thanks{Chengwei Pan is with Institute of Artificial Intelligence, Beihang University, Beijing, China (e-mail: pancw@buaa.edu.cn).}
\thanks{Fandong Zhang is with the AI Lab, Deepwise Healthcare, Beijing, China (e-mail: zhangfandong@deepwise.com).}
\thanks{Xinyang Hu and Xianpeng Wu are with Department of Cardiology of the Second Affiliated Hospital, School of Medicine, Zhejiang University, Hangzhou, China, and Key Laboratory of Cardiovascular of Zhejiang Province, Hangzhou, China (e-mail: hxy0507@zju.edu.cn, wxpzju123@163.com)}
\thanks{G. Zhao, K. Liang and C. Pan have equal contribution.}
}

\maketitle

\begin{abstract}
Vessel segmentation is widely used to help with vascular disease diagnosis. Vessels reconstructed using existing methods are often not sufficiently accurate to meet clinical use standards. This is because 3D vessel structures are highly complicated and exhibit unique characteristics, including sparsity and anisotropy. In this paper, we propose a novel hybrid deep neural network for vessel segmentation. Our network consists of two cascaded subnetworks performing initial and refined segmentation respectively. The second subnetwork further has two tightly coupled components, a traditional CNN-based U-Net and a graph U-Net. Cross-network multi-scale feature fusion is performed between these two U-shaped networks to effectively support high-quality vessel segmentation. The entire cascaded network can be trained from end to end. The graph in the second subnetwork is constructed according to a vessel probability map as well as appearance and semantic similarities in the original CT volume. \red{To tackle the challenges caused by the sparsity and anisotropy of vessels}, a higher percentage of graph nodes are distributed in areas that potentially contain vessels while a higher percentage of edges follow the orientation of potential nearby vessels. Extensive experiments demonstrate our deep network achieves state-of-the-art 3D vessel segmentation performance on multiple public and in-house datasets.
\end{abstract}


\section{Introduction}
\IEEEPARstart{V}{essel} segmentation is widely used in daily practice to characterize many vascular diseases~\cite{grelard2017new,leipsic2014scct}. For example, the obstructed vessels may lead to coronary heart disease, which is the worldwide leading cause of death~\cite{world2008fact,roger2011heart}. Since clinicians mainly rely on interactive tracing and segmentation, vessel reconstruction is traditionally a very time-consuming process and affects the efficiency of diagnosis and intervention. Thus, automatic vessel segmentation can facilitate the reviewing process and plays an important role in medical image analysis.

Over the years, numerous methods have been proposed for automatic vessel segmentation. Due to the state-of-the-art performance of convolutional neural networks (CNNs) on a wide range of pixel-level labelling tasks~\cite{long2015fully,huang2019ccnet,fu2019dual}, CNNs has also been applied to vessel segmentation~\cite{shin2019deep, kong2020learning,wang2020deep}. Nonetheless, the reconstructed vessels are often not sufficiently accurate to meet clinical use standards. This is because vessel structures in 3D CT volumes are highly complicated and exhibit unique characteristics. First, since vessels are thin structures, they only occupy a sparse subset of pixels. Thus, there exists a severe imbalance between vessel and non-vessel pixels. Second, vessel segments are elongated tubular structures that are highly directional and anisotropic.
Conventional CNNs adopt uniform spatial sampling, and therefore, are inept at modeling such sparse and anisotropic structures, giving rise to broken or incomplete results. Thus it becomes critical to design deep neural networks that can effectively exploit the aforementioned characteristics of vessels. 

In this paper, we propose a novel hybrid deep neural network for vessel segmentation. Our network consists of two cascaded subnetworks performing initial and refined segmentation respectively. The second subnetwork further consists of two tightly coupled components, a traditional CNN-based U-shaped network and a graph-based U-shaped network based on graph convolutions. Cross-network multi-scale feature fusion is performed between these two U-shaped networks to effectively support high-quality vessel segmentation. The entire cascaded network can be trained from end to end.

As shown in previous work~\cite{kipf2016semi,velivckovic2017graph,li2019deepgcns}, graph convolutional networks naturally possess a complex shape modeling ability which is well suited for structured data. By setting local regions (supervoxels) in a CT volume as nodes and connections among nearby local regions as edges, the whole CT volume can be regarded as a graph. Specifically, the graph in the second subnetwork is constructed according to a vessel probability map as well as appearance and semantic similarities in the original CT volume. \red{To tackle the challenges brought up by the aforementioned characteristics of vessels}, a higher percentage of graph nodes are distributed in areas that potentially contain vessels while a higher percentage of edges follow the orientation of potential nearby vessels. In addition, the CNN-based U-shaped network is first utilized to extract multi-scale features from the original CT volume. Then at every scale, the features from the CNN are mapped into node features at the corresponding scale of the graph-based U-shaped network and propagated by the GCN at that scale to counteract sparsity and anisotropy. Finally, the enhanced features are reversely mapped into the spatial domain and fused with the original features extracted by the CNN-based U-shaped network.

In summary, our contributions in this paper are three-fold:
\begin{itemize}
\item We propose a cascaded deep neural network for vessel segmentation. The two subnetworks in the cascade are respectively responsible for initial and refined segmentation. There are a pair of tightly coupled U-shaped networks in the second subnetwork of the cascade, one based on CNN and the other based on GCN. Cross-network multi-scale feature fusion is performed between these two U-shaped networks to effectively support high-quality vessel segmentation.
\item We propose a novel way to transform a dense 3D CT volume to a sparse graph format, which can efficiently represent sparse and anisotropic vessel structures. Moreover, our method integrates both appearance and semantic similarities for graph construction.
\item Extensive experiments indicate our deep network achieves state-of-the-art 3D vessel segmentation performance on multiple public and in-house datasets for coronary vessels as well as head and neck vessels, including the public ASOCA dataset.
\end{itemize}

\section{Related Work}
\begin{figure*}[t]
\begin{center}
\includegraphics[width=0.95\linewidth]{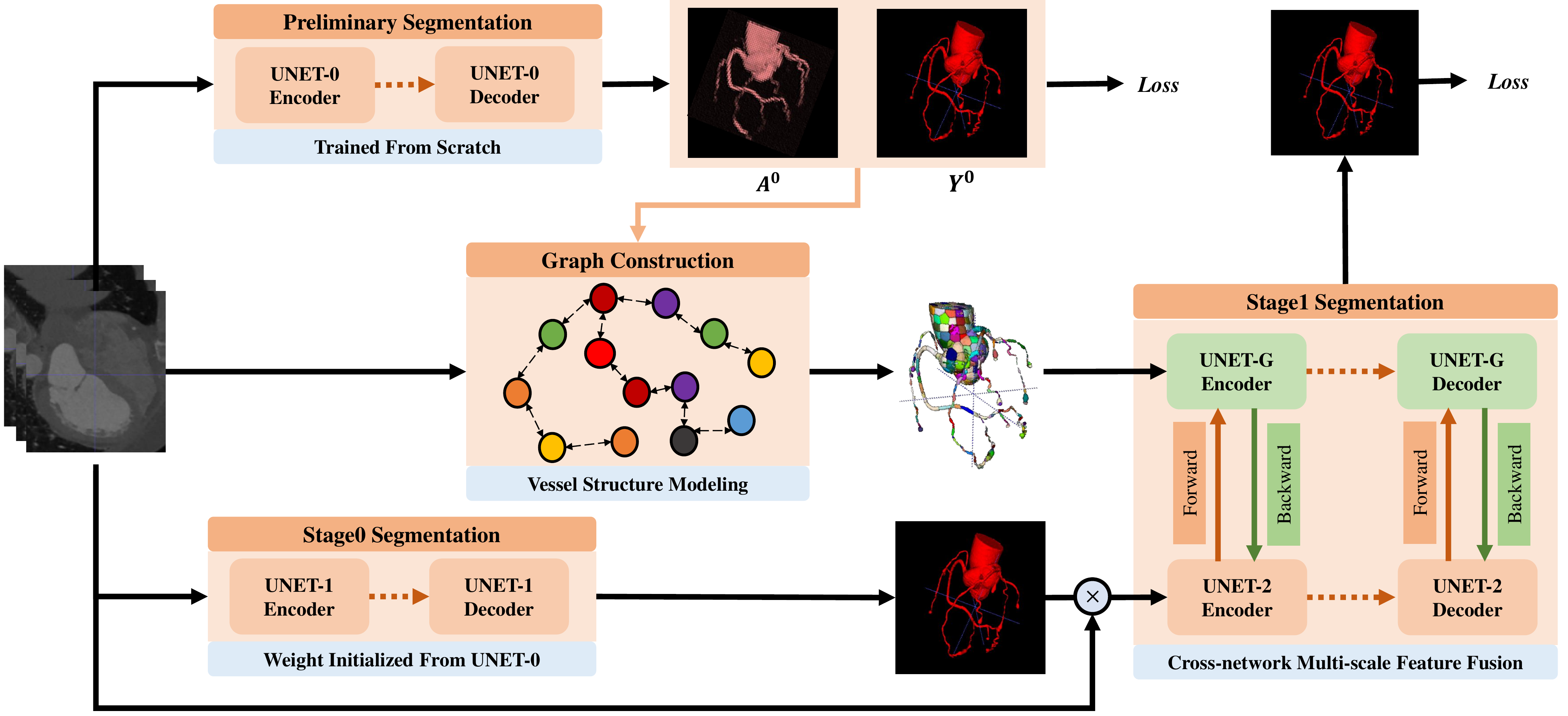}
\end{center}
\caption{Our proposed pipeline for vessel segmentation consists of three stages, preliminary segmentation using a U-Net (UNET-0), graph construction, and final segmentation with a cascaded network, which further consists of two subnetworks with the first subnetwork being a U-Net (UNET-1) and the second subnetwork being a pair of tightly coupled U-shaped networks, a CNN-based U-Net (UNET-2) and a graph U-Net (UNET-G). The preliminary segmentation in the first stage is used by the second stage to construct a graph, whose topology becomes the first level graph in UNET-G.
}\label{fig:frame}
\end{figure*}

\subsection{Graph Convolutional Networks}
Though CNNs achieve impressive performance in many computer vision tasks, they can not efficiently handle graph-structured data. To operate directly on graphs, GCN~\cite{kipf2016semi} is proposed by using layer-wise propagation rule for neural network models. Li et al.~\cite{li2019deepgcns} further adapted the residual/dense connections and dilated convolutions from CNNs into GCN which can solve vanishing gradient problem and increase the depth of GCN. Gao et al.~\cite{gao2019graph} proposed graph pooling and unpooling operations to develop an encoder-decoder model on graph for node classification. The above methods show that GCNs can achieve promising results on modeling graph structure. However, it is still challenging to integrate GCNs into an existing image segmentation framework which is dominated by CNNs.

\subsection{Multi-scale feature modeling}
Multi-scale feature modeling can efficiently capture the global contextual dependencies which plays an important role in image segmentation. Kamnitsas et al.~\cite{kamnitsas2017efficient} proposed a dual pathway deep convolutional neural network. The proposed  dual pathway network incorporates both local and larger contextual information by processing the input images at multiple scales simultaneously. Chen et al.~\cite{chen2017rethinking} proposed to use several parallel atrous convolution with different rates to model the contextual dependencies at multiple scales. Zhao et al.~\cite{zhao2017pyramid} proposed a pyramid pooling module to generate feature maps in different levels for scene parsing. Recently, Tao et al.~\cite{tao2020hierarchical} proposed to combine multi-scale predictions with attention mechanism and achieved the state-of-the-art on Cityscapes and Mapillary Vistas. However, all the above methods adopt uniform spatial sampling for multi-scale feature learning and fail to model the sparsity and anisotropy of vessel.

\subsection{Medical Image Segmentation}
\red{Deep learning has become a methodology of choice for medical image segmentation. Ronneberger et al. proposed UNET~\cite{ronneberger2015u}, which has an encoder-decoder architecture. To avoid missing spatial information, the decoder features from the previous level are up-sampled and combined with the encoder features at the corresponding level through skip connections. The 3D version of UNET~\cite{UNET3d} was further proposed by replacing all 2D operations with their 3D counterparts. In addition, a hybrid densely connected UNET~\cite{li2018h} was proposed to extract intra-slice features with a 2D DenseUNET and aggregate volumetric contexts with its 3D counterpart. Dou et al.~\cite{DOU201740} presented a 3D fully convolutional network equipped with a 3D deep supervision mechanism to combat potential optimization difficulties. Likewise, Zhu et al.~\cite{7965852} proposed to use eight additional deeply supervised layers in their architecture. Jiang et al.~\cite{8417454} developed two multi-resolution residually connected networks to  simultaneously combine features across multiple image resolutions and feature levels. ACSNet~\cite{ACSNet} combines global contexts and local details to deal with the shape and size variations of segmented regions. Similarly, PraNet~\cite{PraNet} aggregates multi-scale features and successively refines the segmentation map through boundary extraction.
Recently, Isensee et al. proposed nnUNET~\cite{isensee2018nnu}, which automatically adapts its architecture according to the geometry of input images. Zhou et al.~\cite{nnFormer} introduced nnFormer, which is an encoder-decoder architecture for volumetric medical image segmentation through the combination of convolution layers and Transformer blocks. 
In addition, the gated axial-attention model in \cite{Valanarasu21} extends the existing architectures and introduces an additional control mechanism with a Local-Global training strategy.
}

\subsection{Vessel Segmentation}
Vessel segmentation plays an important role in medical image analysis. Kong et al.~\cite{kong2020learning} proposed to use a tree-structured convolutional gated recurrent unit (ConvGRU) layer for modeling the anatomical structure of vessels. Since the input of the ConvGRU layer is a uniform local patch, their method cannot well exploit the anisotropy of vessels. Wang et al.~\cite{wang2020deep} proposed a multi-task network to predict a vessel segmentation mask and a distance map. Values in the map represent distances from the center to the surface of every vessel. However, the global structure of vessels is not considered, which limits contextual dependency modeling. There is much work~\cite{shin2019deep,zhang2020graph,yao2020graph,xu2019whole} on the utilization of graph neural networks for vessel segmentation. Shin et al.~\cite{shin2019deep} incorporated a GCN into a CNN architecture to exploit the global structure of vessel shape. However, only the pixel with maximum vessel probability within every rectangular patch is sampled as a graph node, which limits the representation ability of the graph. In addition, GCN features are only calculated at a single scale and do not interact with CNN features. 
In contrast, our framework exhibits a very different way to learn the structural information of vessels. Specifically, we exploit superpixel generation algorithms such as SLIC~\cite{achanta2012slic} to better model the sparsity and anisotropy of vessels, and tightly couple a graph U-Net and a traditional CNN-based U-Net through multi-scale feature fusion across these two networks to better support high-quality vessel segmentation.

\section{Our Framework}
\subsection{Overview}
Consider an input 3D image volume $\textbf{X}\in R^{D \times H \times W}$, where $D$, $H$ and $W$ are the spatial depth, height and width respectively.
The pipeline of our proposed method for vessel segmentation can be decomposed into three stages as shown in Fig. \ref{fig:frame}.

{\flushleft \bf Preliminary Segmentation.} \red{An U-shaped network, UNET-0, is first utilized to create a probability map of the input image volume. This probability map is used for discovering local image regions that have a relatively high probability to contain vessels. Since the probability map may not be very accurate, to reduce the chance of missing regions that actually contain vessels, we apply the dilation operator, a type of image morphological operators, to the probability map to increase the size of image areas with relatively high probability values. The result is a preliminary probability map denoted as $A^0\in (0,1)^{D \times H \times W}$, which is further thresholded to produce a preliminary segmentation mask denoted as $Y^0 \in \{0,1\}^{D \times H \times W}$. The preliminary segmentation mask is used for indicating vessel orientations in regions where vessels are likely to occur. In our experiments, we use a $7 \times 7$ square as the kernel of the dilation operator.}

{\flushleft \bf Graph Construction.} On the basis of the preliminary segmentation mask $Y^0$ and probability map $A^0$, a graph $\mathcal{G} = (\mathcal{V}, \mathcal{E})$ is constructed with a node set $\mathcal{V}$, and an edge set $\mathcal{E}$. To counteract the characteristics of vessel structures including sparsity and anisotropy, a higher percentage of graph nodes are distributed in regions where the preliminary probability map has relatively large values while a higher percentage of edges follow the orientation of the preliminary vessel segmentation mask.

{\flushleft \bf Final Segmentation with a Cascaded Network.} Instead of using a network to refine the preliminary segmentation result obtained in the first stage, we start from scratch and train a cascaded network that takes the original 3D image volume as the input, and performs end-to-end segmentation to produce the final segmentation result. This network consists of two cascaded subnetworks performing initial and refined segmentation respectively. The first subnetwork is an U-shaped network, UNET-1, that shares the same network architecture with UNET-0 in the first stage, but have different network weights because it is trained together with the second subnetwork. The second subnetwork further consists of two tightly coupled components, a traditional CNN-based U-Net (UNET-2) and a graph U-Net (UNET-G)~\cite{gao2019graph}. The graph $\mathcal{G}$ constructed in the second stage becomes the graph with the highest spatial resolution in UNET-G. Cross-network multi-scale feature fusion is performed between UNET-2 and UNET-G to effectively support high-quality vessel segmentation. UNET-1 and UNET-2 are cascaded. The output from UNET-1 includes a hard mask and a soft probability map $P$. Since the input to UNET-2 is the product of $P$ and the original input image $I$, the cascaded network is differentiable. Note that UNET-0 is used to construct graphs as a pre-process. Once the graphs for all training samples have been precomputed, the entire cascaded network can be trained from end to end through gradient backpropagation.

Now let us focus on the second subnetwork. For UNET-2, we represent its convolutional encoder and decoder features as $E_{1:L}^c=\{E_l^c\}$ and $D_{1:L}^c=\{D_l^c\}$ respectively with $L$ being the number of feature levels. The $L^{th}$ decoder and encoder stages have the lowest spatial resolution. UNET-G has the same number of feature levels as UNET-2. The encoder and decoder stages in UNET-2 and UNET-G have one-to-one correspondence. For the $l^{th}$ encoder in UNET-G, its initial graph feature is created as $E_l^g=f(E_l^c, \mathcal{G})$ through a forward mapping function $f(\cdot)$ proposed in ~\cite{liu2020cross} aiming to transform the features between spatial domain and node domain. The forward mapping function $f(\cdot)$ is called KNN-map, which utilizes the K nearest neighborhoods to create the corresponding node feature. Once graph convolutions have been performed on $E_l^g$, the resulting graph convolutional feature is mapped back to the original convolutional feature space of UNET-2 through a backward mapping function $g(\cdot)$ also proposed in ~\cite{liu2020cross} and fused with its initial encoder feature $E_l^c$.

\begin{figure}[t]
\begin{center}
\includegraphics[width=1\linewidth]{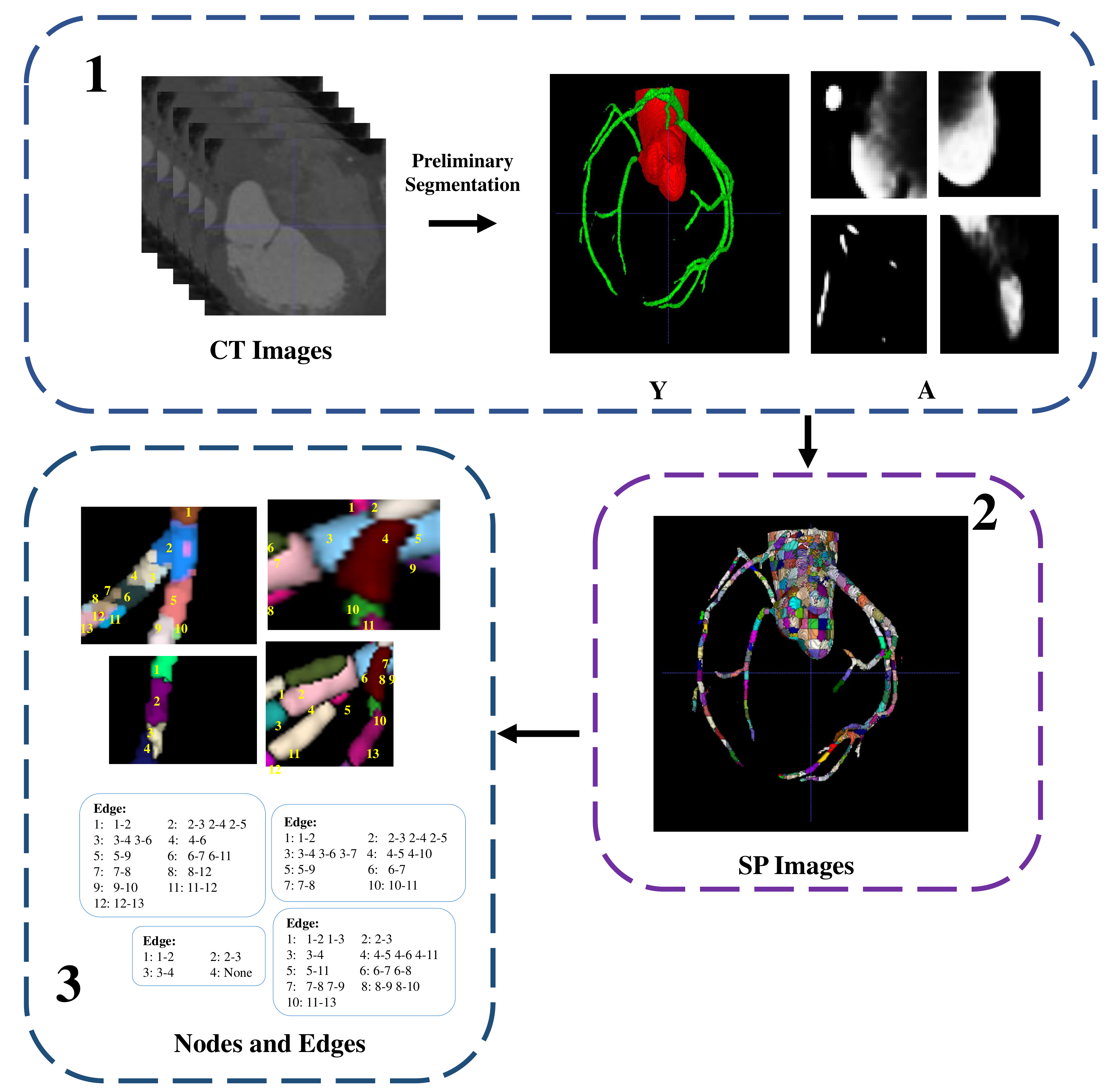}
\end{center}
\caption{A simple example illustrating the graph construction process.}\label{fig:graph-toy}
\end{figure}

\subsection{Graph Construction}
\label{sec:graph}
{\flushleft \bf Graph Nodes.} \red{Since graph neural networks cannot process dense 3D images directly due to high computational cost, we first group all the pixels from a 3D image into super-pixels and then represent each super-pixel as a graph node.}
\red{Here we use the SLIC algorithm~\cite{achanta2012slic} for super-pixel generation.}
In order to capture the 3D structure of vessels, the local region (super-pixel) represented by a graph node should satisfy the following properties: 1) the summation of the vessel probabilities in the region is high; 2) the pixels in the region have similar appearance; 3) the shape of the region follows the local shape of the vessels. \red{The SLIC algorithm is based on a distance measure, which originally consists of two terms, grayscale difference and Euclidean distance. To satisfy the aforementioned properties, we add a third term based on geodesic distance. The updated distance measure for SLIC and its three terms are formulated as follows.}
\begin{equation}
\begin{aligned}
d(i, j) = d_{gray}(i,j) + d_{dis}(i,j) + d_{geo}(i,j),
\label{dis}
\end{aligned}
\end{equation}
where
\begin{equation}
\begin{aligned}
d_{gray}(i,j) = |X_{i} - X_{j}|,
\label{semdis}
\end{aligned}
\end{equation}
\begin{equation}
\begin{aligned}
d_{dis}(i,j) = \sqrt{(x_{i} - x_{j})^{2} + (y_{i} - y_{j})^{2} + (z_{i} - z_{j})^{2}},
\label{disdis}
\end{aligned}
\end{equation}
\begin{equation}
\begin{aligned}
d_{geo}(i, j) = \min_{\mathcal{Q} \in \mathcal{P}_{i,j}} \sum_{q\in \mathcal{Q}}  A_q^{0} \| \nabla(X_q + X_q^{0})\cdot u_q \|,
\label{geodis}
\end{aligned}
\end{equation}
where $X_i$ denotes the gray scale of the $i^{th}$ pixel, $[x_i, y_i, z_i]^{T}$ denotes its spatial coordinates, $\mathcal{P}_{i,j}$ represents the complete set of paths from pixel $i$ to pixel $j$, $\mathcal{Q}$ denotes one path in $\mathcal{P}_{i,j}$, $q$ denotes any pixel on $\mathcal{Q}$, and $u_q$ represents the unit tangent vector of path $\mathcal{Q}$ at $q$. The geodesic distance between two points is defined as the minimum of the integration of $X$, $X^{0}$ and $A^{0}$ as in (\ref{geodis}) among all the paths in $\mathcal{P}_{i,j}$, where $X^{0} = X\circ Y^{0}$ and $\circ$ stands for element-wise multiplication.
$\nabla(X_q + X_q^0)$ means the gradient of $X_q + X_q^0$.
$X_q + X_q^0$ doubles the value in vessel areas, therefore, it will create more graph nodes in vessels because of a larger distance between different nodes in these areas.
In practice, we use the Dijkstra's algorithm to calculate the geodesic distance. The definition of geodesic distance in (\ref{geodis}) ensures that regions potentially containing vessels have a higher density of graph nodes. Note that the three distance terms have been individually normalized before added together in the overall distance measure.

{\flushleft \bf Graph Edges.}
As there typically exist a large number of graph nodes in a 3D image volume, in this paper, we only consider locally connected graphs to reduce computational cost. Each node $i$ is only connected to other nearby nodes whose geodesic distance is below a predefined threshold $t_{geo}$. That is, there exists an edge between nodes $i$ and $j$ if and only if $d'_{geo}(i, j) < t_{geo}$, where $d'_{geo}(i, j)$ is a modified version of the geodesic distance in (\ref{geodis}) where $A^{0}$ is replaced with $(1-A^{0})$. Since our geodesic distance is affected by the vessel mask and probability map, this connection rule implies that the Euclidean distance between two connected nodes has a larger threshold when the nodes are near potential vessels and the orientation of the edge between the nodes roughly follows the local orientation of the preliminary vessel mask. As a result, the constructed graph has denser and longer connections in regions potentially containing vessels.

In our constructed graph, every edge is associated with an edge weight, which is a product of two components, $e_w= e_w^{s}e_w^{a}$,
where $e_w^s$ and $e_w^a$ represent semantic consistency and appearance similarity respectively.
For a convolutional feature map $F$ in UNET-2, we first create its node representation $F_{V} \in \mathcal{R}^{|V| \times C}$ through the forward mapping function $f(\cdot)$ in \cite{liu2020cross} on the feature map $F$. Then we define the semantic consistency of the edge between nodes $i$ and $j$ as
\begin{equation}
\begin{aligned}
e_w^{s}(i, j) = \sigma([F_{V}^{i}, F_{V}^{j}]w^{s}),
\label{eq:sedge}
\end{aligned}
\end{equation}
where $F_{V}^{i}, F_{V}^{j}$ represent the $i$-th and $j$-th node features, $w^{s} \in \mathcal{R}^{2C}$ is a trainable weight vector fusing the two node features, and $\sigma(\cdot)$ is the sigmoid activation function. [] means a concatenation.

We use the gray-scale information associated with graph nodes to define the appearance similarity of an edge as
\begin{equation}
\begin{aligned}
e_w^{a}(i, j) = \sigma([F_{X}^{i}, F_{X}^{j}]w^{g})
\label{eq:gedge}
\end{aligned}
\end{equation}
where $F_{X} = f(X \circ Y^0 \circ A^0, \mathcal{G})$, $w_{g} \in \mathcal{R}^{2C}$ is another trainable weight vector fusing the mapped features at the two nodes. Instead of using the gray-scale information $X$ from the input image volume only, we also include the semantic information $Y^0$ and $A^0$ from the preliminary segmentation to focus on potential vessel regions.

A simple example illustrating the above graph construction process is given in Fig.~\ref{fig:graph-toy}.

\begin{table}[!tb]
\begin{center}
\begin{tabular}{l c l}
\hline
Layers & Output size & UNET-0,1,2\\
\hline
\hline
en-conv0    & $256 \times 256 \times 256$ & conv($3 \times 3 \times 3$, 16) \\
en-conv1    & $128 \times 128 \times 128$&2 $\times$ BuildBlock($3 \times 3 \times 3$, 32) \\
en-conv2	& $64 \times 64 \times 64$  & 2 $\times$ BuildBlock($3 \times 3 \times 3$, 64)) \\
en-conv3	& $32 \times 32 \times 32$  & 2 $\times$ BuildBlock($3 \times 3 \times 3$, 128) \\
en-conv4	& $16 \times 16 \times 16$  & 2 $\times$ BuildBlock($3 \times 3 \times 3$, 256) \\
de-conv4	& $16 \times 16 \times 16$  & 2 $\times$ BuildBlock($3 \times 3 \times 3$, 256) \\
de-conv3    & $32 \times 32 \times 32$ &  2 $\times$ BuildBlock($3 \times 3 \times 3$, 128)  \\
de-conv2	& $64 \times 64 \times 64$  & 2 $\times$ BuildBlock($3 \times 3 \times 3$, 64) \\
de-conv1	& $128 \times 128 \times 128$&2 $\times$ BuildBlock($3 \times 3 \times 3$, 32) \\
de-conv0    & $256 \times 256 \times 256$&2 $\times$ BuildBlock($3 \times 3 \times 3$, 16) \\
classifier	& $256 \times 256 \times 256$& conv($1 \times 1 \times 1$, 2)  \\
\hline
\end{tabular}
\end{center}
\caption{Network architecture of UNET-0,1,2 used in the proposed pipeline. Convolution layers in the encoder of the original U-Net are replaced with residual blocks. Inside the brackets are the shape of the residual blocks, and outside the brackets is the number of stacked blocks in a stage. Downsampling (max pooling) is performed after en-conv0, en-conv1, en-conv2, en-conv3 with stride 2, respectively.  Upsampling is performed after each de-conv stage, and the number of input channels of each layer can be found from the preceding layer.}\label{tab:network}
\end{table}

\begin{figure*}[t]
\begin{center}
\includegraphics[width=\linewidth]{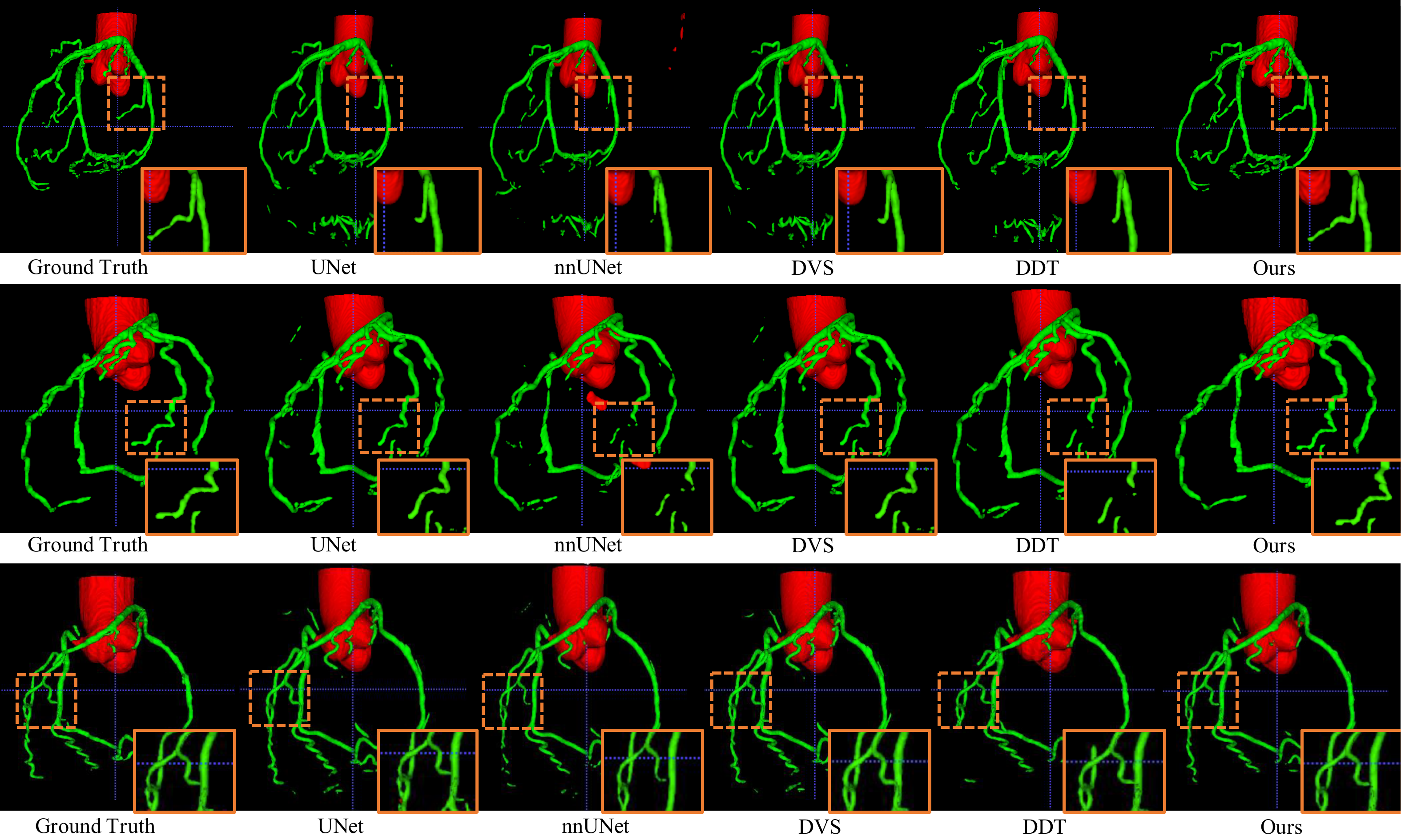}
\end{center}
\caption{From left to right, it is the ground truth, the results of UNET, nnUNET, DVS, DDT and our model, respectively. The aorta and the coronary vessels are marked with red and green. Although DDT achieves the best performance compared with other previous state-of-the-art methods, it may generate incomplete vessel masks when the structure of vessels is complicated.}\label{fig:visual}
\end{figure*}

\begin{figure*}[t]
\begin{center}
\includegraphics[width=\linewidth]{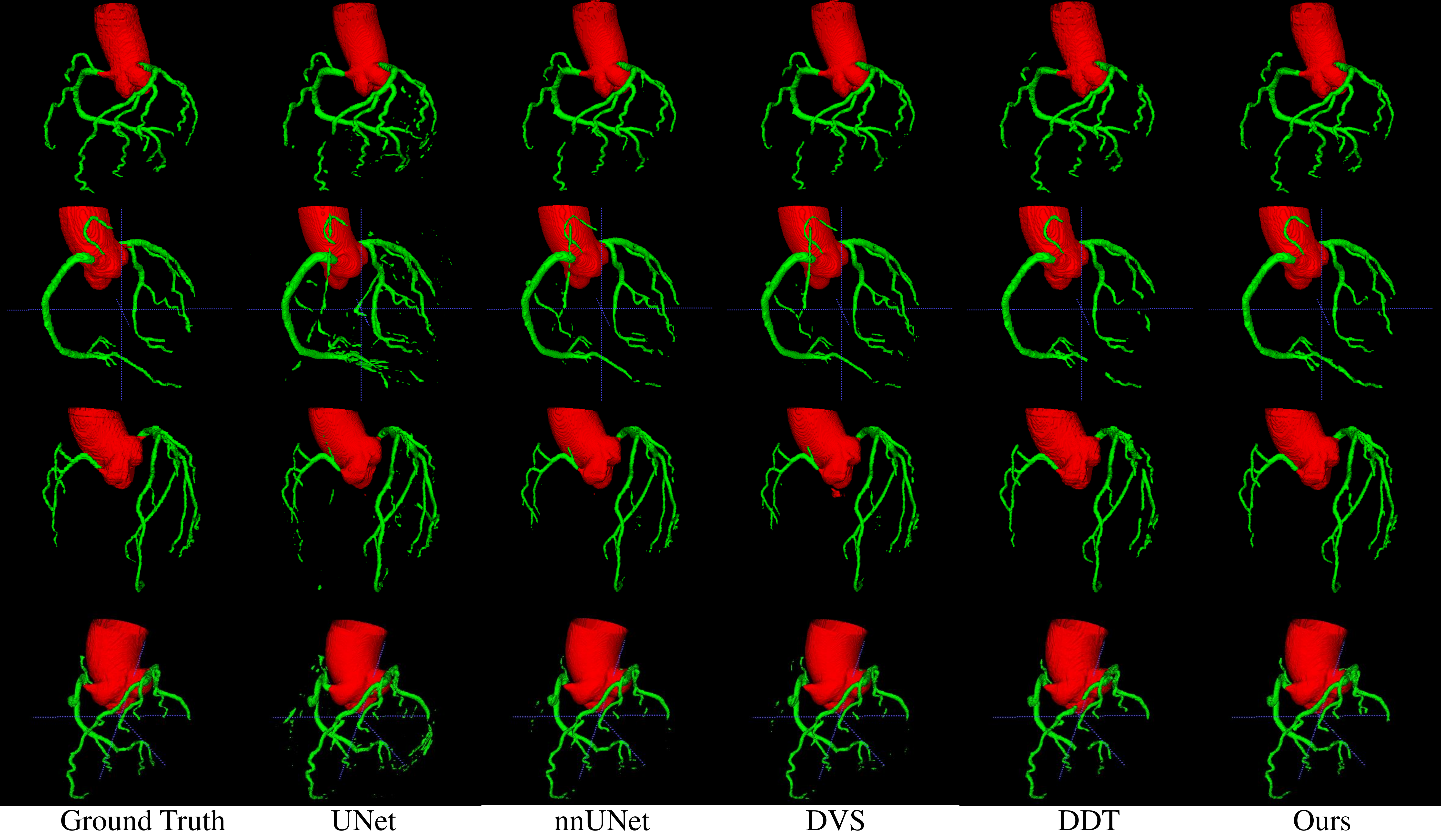}
\end{center}
\caption{Sample visual results on the ACA dataset. From left to right, it is the ground truth, the results of UNET, nnUNET, DVS, DDT and our model, respectively. The aorta and the coronary vessels are marked with red and green. Although DDT achieves the best performance compared with other previous state-of-the-art methods, it may generate incomplete vessel masks when the structure of vessels is complicated.}\label{fig:visual0}
\end{figure*}

\subsection{Cross-Network Multi-Scale Feature Fusion}
The features from UNET-2 need to be mapped into the node domain of UNET-G and further enhanced through graph convolutions over the constructed graph structure to better observe global priors of vessel connectivity. Afterwards we reversely map the enhanced features to the spatial domain of UNET-2 and fuse them with the original features there through a residual connection.

{\flushleft \bf Encoder Feature Fusion.} The encoder feature map $E_l^c$ from the $l^{th}$ stage of UNET-2 is transformed into node features at the corresponding level of UNET-G through the forward mapping function $f(\cdot)$ defined in \cite{liu2020cross}. Then the mapped features are fused with the down-sampled encoder features from the previous stage in UNET-G. A residual graph convolution module $\Omega( \cdot)$ is utilized to enhance the fused features for more accurately modeling complex vessel structures and better observing global priors of vessel connectivity. Therefore, the graph convolutional encoder features at the $l^{th}$ stage of UNET-G are created as
\begin{equation}
\begin{aligned}
E_l^g=\Omega(f(E_l^c, \mathcal{G})+\mbox{down}(E_{l-1}^g)).
\end{aligned}\label{eq:gefeature}
\end{equation}
Then the graph convolutional features $E_l^g$ are reversely mapped to the original convolutional feature space of UNET-2 through the backward mapping function $g(\cdot)$ defined in \cite{liu2020cross} and fused with its initial encoder feature to produce the enhanced encoder feature at the $l^{th}$ level,
\begin{equation}
\begin{aligned}
E_{l}=g(E_l^g) + E_l^c.
\end{aligned}\label{eq:cefeature}
\end{equation}

\begin{table}[!tb]
\begin{center}
\begin{tabular}{c c c}
\hline
Dataset & Avg \#Nodes per Image & Avg \#Edges per Node\\
\hline
\hline
ASOCA    &  12000   &  8.12     \\
ACA      &  9600    &  7.11     \\
HNA      &  13000   &  7.03     \\
\hline
\end{tabular}
\end{center}
\caption{Statistics of constructed graphs. Average number of nodes per image is calculated using all images in a dataset. And average number of edges per node is calculated using all nodes in a dataset.}\label{tab:edge}
\end{table}

\begin{table}[!tb]
\begin{center}
\begin{tabular}{c c c}
\hline
Dataset & Avg \#Nodes per Image & Avg \#Edges per Node\\
\hline
\hline
\multicolumn{3}{c}{Set1\_1: n\_segmetns is 28000} \\
\hline
ASOCA    &  19010   &  8.67     \\
ACA      &  14300    &  7.40     \\
HNA      &  22420   &  8.10    \\
\hline
\multicolumn{3}{c}{Set1\_2: n\_segmetns is 14000} \\
\hline
ASOCA    &  12000   &  8.12     \\
ACA      &  9600    &  7.11     \\
HNA      &  13000   &  7.03     \\
\hline
\hline
\multicolumn{3}{c}{Set1\_3: n\_segmetns is 7000} \\
\hline
ASOCA    &  6020   &  6.12     \\
ACA      &  3110    &  4.11     \\
HNA      &  5200   &  4.03     \\
\hline
\hline
\multicolumn{3}{c}{Set1\_4: n\_segmetns is 3500} \\
\hline
ASOCA    &  3210   &  4.12     \\
ACA      &  2930    &  3.03     \\
HNA      &  2122   &  2.14     \\
\hline
\end{tabular}
\end{center}
\caption{Average number of graph nodes and edges for different values of n\_segments when min\_size\_factor is fixed to 0.5.}\label{tab:abl_edge_segment}
\end{table}

\begin{table}[!tb]
\begin{center}
\begin{tabular}{c c c}
\hline
Dataset & Avg \#Nodes per Image & Avg \#Edges per Node\\
\hline
\hline
\multicolumn{3}{c}{Set2\_1: min\_size\_factor is 0.3} \\
\hline
ASOCA    &  8900   &  8.01     \\
ACA      &  7230    &  6.89     \\
HNA      &  9600   &  7.01     \\
\hline
\hline
\multicolumn{3}{c}{Set2\_2: min\_size\_factor is 0.4} \\
\hline
ASOCA    &  10300   &  8.02     \\
ACA      &  8410    &  7.12     \\
HNA      &  11200   &  7.13     \\
\hline
\hline
\multicolumn{3}{c}{Set2\_3: min\_size\_factor is 0.5} \\
\hline
ASOCA    &  12000   &  8.12     \\
ACA      &  9600    &  7.11     \\
HNA      &  13000   &  7.03     \\
\hline
\hline
\multicolumn{3}{c}{Set2\_4 min\_size\_factor is 0.6} \\
\hline
ASOCA    &  12100   &  8.13     \\
ACA      &  9870    &  7.21     \\
HNA      &  13210   &  7.13     \\
\hline
\end{tabular}
\end{center}
\caption{Average number of graph nodes and edges for different values of min\_size\_factor. n\_segments is fixed to 14000.}\label{tab:abl_edge_size}
\end{table}

{\flushleft \bf Decoder Feature Fusion.} The decoder feature $D_l^c$ from the $l^{th}$ stage of UNET-2 is transformed into node features at the corresponding level of UNET-G through the same forward mapping function $f(\cdot)$. Then the mapped features are fused with the up-sampled decoder features from the previous stage in UNET-G, and the fused features are enhanced with the same residual graph convolution module $\Omega( \cdot)$ before further fused with the graph encoder feature $E_l^g$ through the skip connection at the $l^{th}$ stage of UNET-G. Thus the graph convolutional decoder features at the $l^{th}$ stage of UNET-G are defined as
\begin{equation}
\begin{aligned}
D_l^g=\Omega(f(D_l^c, \mathcal{G})+\mbox{up}(D_{l+1}^g)) + E_{l}^{g}.
\end{aligned}\label{eq:gdfeature}
\end{equation}
Then the graph convolutional decoder features $D_l^g$ are reversely mapped to the original feature space of UNET-2 through the same backward mapping function $g(\cdot)$. We further fuse the reversely mapped features with both the initial decoder feature of UNET-2 and the skip-connected enhanced encoder feature $E_l$ to produce the enhanced decoder feature at the $l^{th}$ level,
\begin{equation}
\begin{aligned}
D_l=g(D_l^g) + D_l^c + E_{l}.
\end{aligned}\label{eq:cdfeature}
\end{equation}
The last enhanced decoder feature is used to produce the final segmentation of vessels with a pixel-wise softmax classifier.

\red{
{\flushleft \bf Forward and Backward Mappings} 
We adopt the forward and backward mapping functions defined in \cite{liu2020cross} to map pixel-level features in a CNN-based U-Net to node features in a graph U-Net and vice versa. The key consideration during feature mapping design lies in revealing the relations between node and pixel-level features. As illustrated in the following equations, the kNN (k Nearest Neighbor) based forward mapping $\phi_k$ with its auxiliary matrix $A$ aggregates pixel-level features over irregular regions to obtain corresponding node features adaptively according to their spatial relations.
\begin{equation}
\begin{aligned}
	\phi_k(F, \mathcal{N}) = (Q^f)^{T} F,
\end{aligned}
\end{equation}

\begin{equation}
\begin{aligned}
	Q^{f} = A (\Lambda^{f})^{-1},
\end{aligned}
\end{equation}
\begin{equation}
\label{assign_matrix}
\begin{aligned}
	A_{ij}  =
		\begin{cases}
		1 & \text{if $j$ th node is kNN of $i$ th pixel} \\
	    0 & \text{Otherwise}
		\end{cases},
\end{aligned}
\end{equation}
where $\mathcal{N} \in \{\mathcal{V}, \mathcal{U} \}$ represents the node set corresponding to pixel-level spatial visual features $F \in \mathbb{R} ^ {HW \times C}$, $A \in \mathbb{R} ^ {HW \times |\mathcal{N} |}$  is an auxiliary matrix that assigns spatial features to k nearest graph nodes, $\Lambda^{f} \in \mathbb{R}^{\mathcal{|\mathcal{N}| \times |\mathcal{N}|}} $ is a diagonal matrix,  $\Lambda^{f}_{jj} = \sum\limits_{i=1}^{HW} A_{ij}$, and $Q ^{f} \in \mathbb{R} ^ {HW \times |\mathcal{N}|}$ is a normalized form of $A$ and serves as the forward mapping matrix.

The backward mapping function $\psi_{k}$ projects each graph node feature back to the spatial domain. The backward mapping follows similar design principles as the forward mapping and makes use of the same number of nearest neighbors. Formally, $\psi_{k}$ is formulated as follows.
\begin{equation}
\begin{aligned}
	\psi_k(Z, \mathcal{N}) = Q^r [Z]_e,
\end{aligned}
\end{equation}
\begin{equation}
\begin{aligned}
	Q^{r} = (\Lambda^{r})^{-1} A,
\end{aligned}
\end{equation}
where $\mathcal{N} \in \{\mathcal{V}, \mathcal{U}\}$ represents the node set of the graph,  $A \in \mathbb{R} ^ {HW \times |\mathcal{N} |}$ is similar to the definition in Equation \ref{assign_matrix}, $[\cdot]_e$ indicates the indexing operator which selects nodes in the graph, $\Lambda^{r} \in \mathbb{R}^{HW \times HW} $ is a diagonal matrix, $\Lambda^{r}_{ii} = \sum\limits_{j=1}^{|\mathcal{N}|} A_{ij}$, and $Q ^{r} \in \mathbb{R} ^ {HW \times |\mathcal{N}|}$ is the backward mapping matrix, which is also a normalized form of $A$.
}

\begin{figure*}[t]
\begin{center}
\includegraphics[width=1\linewidth]{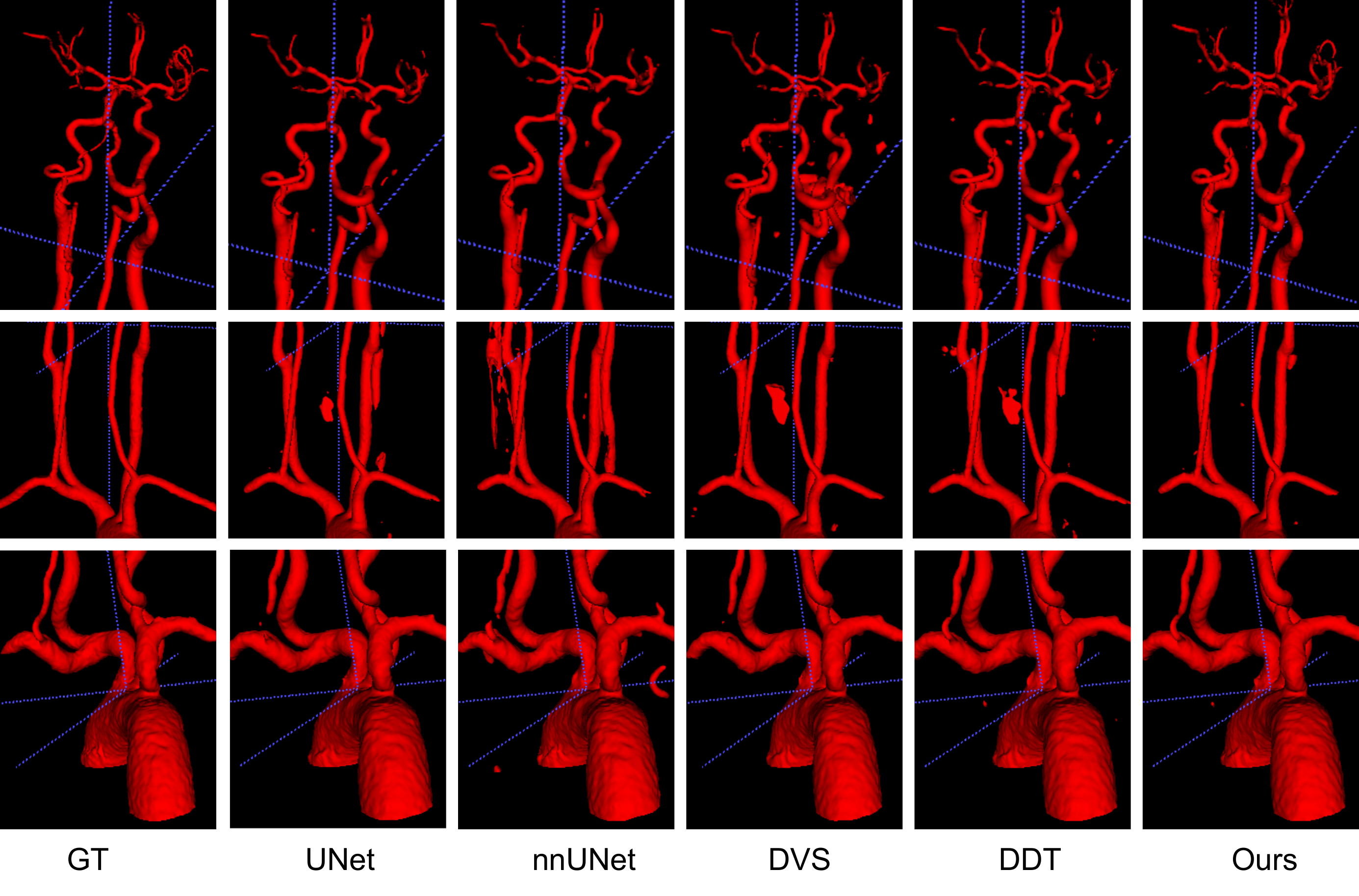}
\end{center}
\caption{Sample visual results on the HNA dataset. From left to right, it is the ground truth, the results of UNET, nnUNET, DVS, DDT and our model, respectively.}\label{fig:visual3}
\end{figure*}
\begin{figure*}[t]
\begin{center}
\includegraphics[width=1\linewidth]{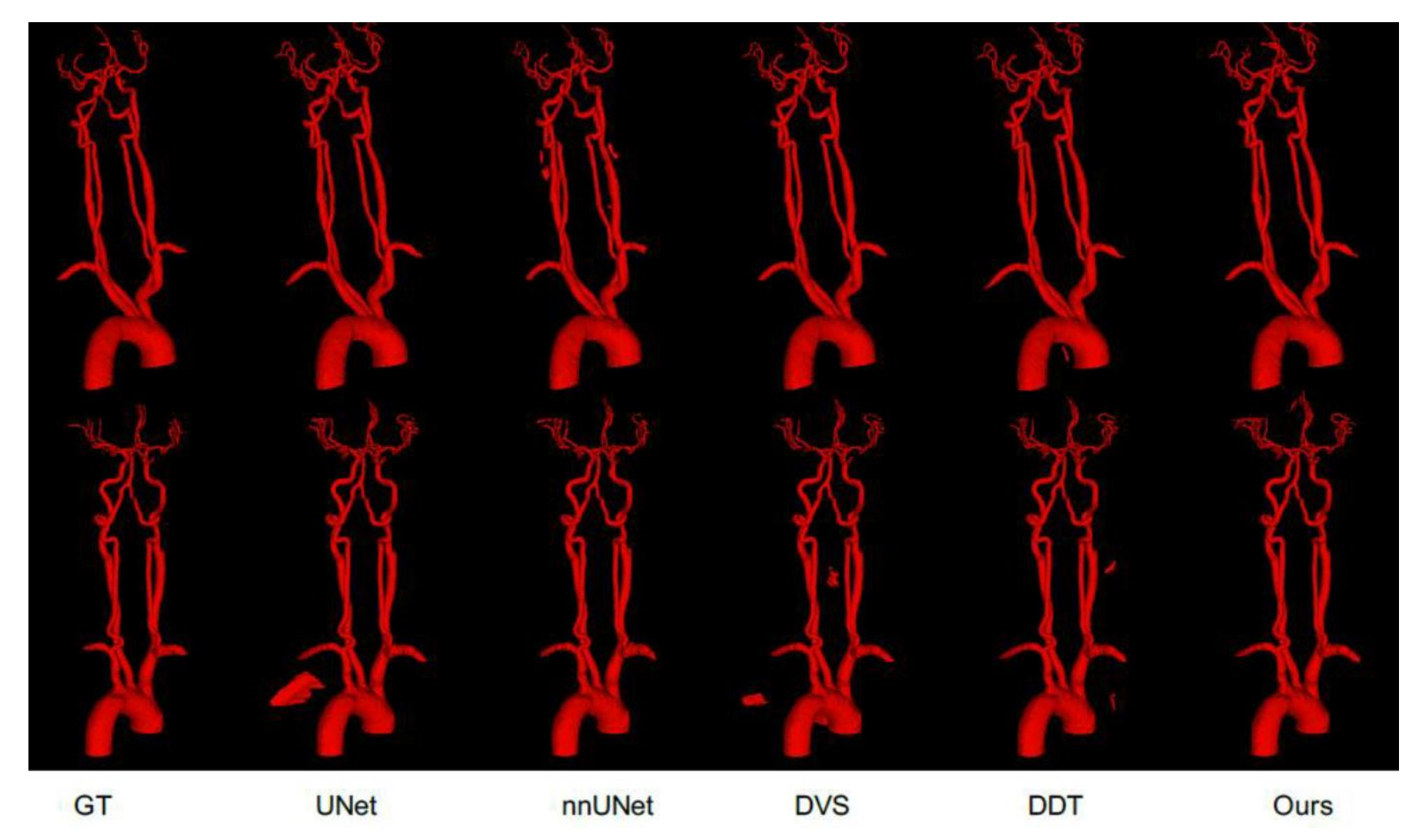}
\end{center}
\caption{Sample visual results on the HNA dataset. From left to right are the ground truth, the results of UNET, nnUNET, DVS, DDT and our model, respectively.}\label{fig:visual3}
\end{figure*}

\section{Experiments}

\subsection{Datasets}
\black{{\noindent \bf ASOCA} Automated Segmentation of Coronary Arteries Dataset (ASOCA) is a public dataset in MICCAI-2020 challenge \footnote{https://asoca.grand-challenge.org} which aims to segment the coronary artery lumen. The dataset consists of a training set of 40 Cardiac Computed Tomography Angiography (CCTA) images and a test set of 20 CCTA images. The images in the testing set were annotated and verified by experts we invited.} \red{The original image resolution of the ASOCA dataset is 512$\times$512$\times$$N$, where $N$ is between 168 and 334. 
}

\black{{\noindent \bf ACA} Aorta and Coronary Artery Dataset (ACA) is an in-house dataset which contains 1000 CCTA images. The dataset is utilized to segment both aorta and coronary arteries. Each image is annotated by one expert annotator and verified by a second expert. We split the dataset into a training set of 800 images, a validation set of 100 images and a test set of 100 images.} \red{The original image resolution of the ACA dataset is 512$\times$512$\times$$N$, where $N$ is between 192 and 600. 
}

\black{{\noindent \bf HNA} Head and Neck Artery Dataset (HNA) is an in-house dataset which contains 800 CT angiography (CTA) images of head and neck. The images are annotated in the same way as ACA. Cerebral, vertebral and carotid arteries are annotated as the target vessel mask. The dataset is split into a training set of 640 images, a validation set of 80 images and a test set of 80 images.}
\red{The original image resolution of the HNA dataset is also 512$\times$512$\times$$N$, where $N$ is between 192 and 600. 
}

\subsection{Experimental Setup}
\black{{\flushleft \bf Evaluation Metrics.} Dice coefficient (DICE) and average symmetric surface distance (ASSD) (ASSD is measured in millimeters) are adopted as the evaluation metrics since they are commonly used in medical image segmentation \cite{yue2019cardiac}. In addition, to evaluate unique characteristics of tubular structure, another two metrics called skeleton recall (SR) and skeleton precision (SP) are defined as follows:
\begin{equation}
\begin{aligned}
SR(S, G)=\frac{|S\bigcap Q(G)|}{|Q(G)|},
\end{aligned}\label{eq:sr}
\end{equation}

\begin{equation}
\begin{aligned}
SP(S, G)=\frac{|Q(S)\bigcap G|}{|Q(S)|},
\end{aligned}\label{eq:sp}
\end{equation}
where $S$ and $G$ are the segmentation result and the ground truth annotation respectively. The function $Q(\cdot)$ is used to acquire the skeleton of a tubular mask, which can preserve original vascular topology and connectivity. Here we use skeletonization function \cite{scikit-image} as the implementation of $Q(\cdot)$.

{\flushleft \bf Network Structure and Training.} Each sub-network of the proposed method is a U-shaped network. All CNN-based U-shaped networks, including UNET-0, UNET-1 and UNET-2 are based on the original U-Net~\cite{ronneberger2015u} except that the original convolution layers in its encoder are replaced with residual blocks~\cite{he2016deep}. The network architecture of UNET-0,1,2 used in the proposed pipeline is given in Table~\ref{tab:network}. UNET-G is a graph U-Net~\cite{gao2019graph}. Each downsampling operation in UNET-G halves the number of graph nodes, and each upsampling operation doubles the number of nodes. The feature dimension of every graph node is always set to 256 in all the experiments reported in this paper. The input image is always resized to $256 \times 256 \times 256$, and the batch size on a single GPU is 2.} \red{The proposed cascaded network is trained by jointly optimizing the weighted cross-entropy loss, $\mathcal{L}_{wbce}=-\beta y\cdot\log(p)-(1-y)\cdot\log(1-p)$, and the dice loss, $\mathcal{L}_{Dice}=1-\frac{2y\cdot p}{\|y\|_1+\|p\|_1}$, where $y$ and $p$ are the ground-truth and predicted masks, respectively. We set $\beta=5$ to increase the vessel recall.} \black{All models are trained for 100 epochs from scratch using PyTorch~\cite{paszke2019pytorch} on NVIDIA Titan Xp pascal GPUs. We set the weight decay to 1e-4 and use Adam~\cite{kingma2014adam} as the optimizer with the initial learning rate set to 1e-4. The learning rate is reduced by a factor of 10 after every 40 epochs.}

\black{{\flushleft \bf Graph Hyperparameter Setting.}
We use a 3D version of the SLIC algorithm~\cite{achanta2012slic} to generate superpixels. Two parameters of the algorithm control the total number of superpixels in an image. One of them is `n\_segments', which is the maximum} number of superpixels, and the other is `min\_size\_factor', which defines the ratio between the minimum size of a superpixel and the average size of a superpixel. In the experiments reported in this paper, `n\_segments' is always set to 14000, and `min\_size\_factor' is set between 0.3 and 0.65.

In a graph, each node $i$ is only connected to other nearby nodes whose geodesic distance is below a predefined threshold $t_{geo}$. That is, there exists an edge between nodes $i$ and $j$ if and only if $d'_{geo}(i, j) < t_{geo}$, where $d'_{geo}(i, j)$ is the geodesic distance between nodes $i$ and $j$. $t_{geo}$ is a hyperparameter that needs to be empirically set only once for each dataset. For the ASOCA dataset, $t_{geo}$ is set to 0.30. For the ACA dataset, $t_{geo}$ is set to 0.35. For the HNA dataset, $t_{geo}$ is set to 0.40. 

Table~\ref{tab:edge} shows the statistics of graph nodes and edges.

\begin{table}[t]
\begin{center}
\begin{tabular}{l c c c c}
\hline
Method & DICE (\%) & ASSD  & SP (\%) & SR (\%)\\
\hline
\hline
	LB-1st & 88.56 & - & - & -\\
	LB-2nd & 88.00 & - & - & -\\
    LB-3rd &  87.94 & - & - & -\\
    LB-4th	& 87.36 & - & - & -\\
    LB-5th	& 87.17 & - & - & -\\
    LB-6th	& 87.11 & - & - & -\\
\hline
    DDT \cite{wang2020deep} & 88.21  & 0.571 & 95.0 & 94.5 \\
	DVS \cite{shin2019deep}  & 87.32 & 0.582 & 94.1 & 93.2  \\
    UNET3d \cite{UNET3d} & 83.20 & 0.644 & 93.0 & 91.0  \\
    \red{ResUNET} \cite{zhang2018road} & 83.20 & 0.644 & 93.0 & 91.0  \\
    \red{DenseUNET} \cite{li2018h} & 83.20 & 0.644 & 93.0 & 91.0  \\
    \red{H-DenseUNET} \cite{li2018h} & 83.20 & 0.644 & 93.0 & 91.0  \\
    nnUNET \cite{isensee2018nnu} & 85.11 & 0.572 & 94.3 & 92.4  \\
	PSP-Net \cite{zhao2017pyramid}  & 84.12 & 0.593 & 94.2 & 92.1   \\
	HMSA \cite{tao2020hierarchical}  & 86.23 & 0.561 & 95.2 & 93.3   \\
    \red{Ours} & \textbf{89.89} & \textbf{0.544} & \textbf{95.6} & \textbf{95.9}  \\
    \red{Ours+ResUNET} & \textbf{89.90} & \textbf{0.541} & \textbf{95.7} & \textbf{95.9}  \\
    \red{Ours+DenseUNET} & \textbf{89.89} & \textbf{0.540} & \textbf{95.7} & \textbf{95.9}  \\
    \red{Ours+H-DenseUNET} & \textbf{89.91} & \textbf{0.530} & \textbf{95.8} & \textbf{96.0}  \\
\hline
\end{tabular}
\end{center}
\caption{Performance comparison on the ASOCA Dataset among state-of-the-art segmentation algorithms. The results of MICCAI Leaderboard are shown in \url{https://asoca.grand-challenge.org/evaluation/challenge/leaderboard/}, which only shows the performance in terms of DICE. For other methods, we evaluate them in terms of four performance metrics including DICE, ASSD, SP and SR.}\label{tab:miccaisota}
\end{table}

\begin{table}[t]
\begin{center}
\begin{tabular}{ll c c c c}
\hline
Method & DICE (\%) & ASSD & SP (\%) & SR (\%)\\
\hline
\hline
    DDT \cite{wang2020deep} & 91.2 & 0.497 & 96.0 & 89.2\\
	DVS \cite{shin2019deep}  & 90.1 & 0.503 & 95.1 & 88.3  \\
    UNET3d \cite{UNET3d} & 87.3 & 0.711 & 94.0 & 89.4 \\
    \red{ResUNET} \cite{zhang2018road}  & 88.4 & 0.612 & 95.1 & 89.6 \\
    \red{DenseUNET} \cite{li2018h} & 88.9 & 0.568 & 95.2 & 89.4 \\
    \red{H-DenseUNET} \cite{li2018h} & 89.9 & 0.528 & 95.3 & 90.1 \\
    nnUNET \cite{isensee2018nnu} & 88.3 & 0.630 & 95.5 & 90.6  \\
	PSP-Net \cite{zhao2017pyramid}  & 89.0 & 0.642 & 95.0 & 89.6  \\
	HMSA \cite{tao2020hierarchical} & 90.2 & 0.592 & 96.7 & 90.1  \\
	\hline
    \red{Ours} & \textbf{94.2} & \textbf{0.448} & \textbf{97.1} & \textbf{95.1}  \\
    \red{Ours+ResUNET} & \textbf{94.6} & \textbf{0.445} & \textbf{97.3} & \textbf{95.2}  \\
    \red{Ours+DenseUNET} & \textbf{94.8} & \textbf{0.444} & \textbf{97.3} & \textbf{95.3}  \\
    \red{Ours+H-DenseUNET} & \textbf{94.8} & \textbf{0.443} & \textbf{97.3} & \textbf{95.2}  \\
\hline
\end{tabular}
\end{center}
\caption{Performance comparison on the ACA dataset among state-of-the-art segmentation algorithms.}\label{tab:vhousesota}
\begin{center}
\begin{tabular}{ll c c c c}
\hline
Method & DICE (\%) & ASSD & SP (\%) & SR (\%)\\
\hline
\hline
    DDT \cite{wang2020deep} & 92.4 & 0.401 & 96.1 & 93.3\\
	DVS \cite{shin2019deep}  & 91.3 & 0.472 & 97.2 & 92.4  \\
    UNET3d \cite{UNET3d} & 87.3 & 0.664 & 95.1 & 90.5 \\    
    \red{ResUNET} \cite{zhang2018road} & 87.7 & 0.661 & 95.2 & 90.7 \\
    \red{DenseUNET} \cite{li2018h} & 88.1 & 0.618 & 95.3 & 90.8 \\
    \red{H-DenseUNET} \cite{li2018h} & 89.1 & 0.588 & 95.7 & 92.0 \\
    nnUNET \cite{isensee2018nnu} & 89.9 & 0.600 & 94.3 & 91.2  \\
	PSP-Net \cite{zhao2017pyramid}  & 90.1 & 0.593 & 94.7 & 90.0  \\
	HMSA \cite{tao2020hierarchical}  & 91.4 & 0.543 & 95.6 & 91.1  \\
	\hline
    \red{Ours} & \textbf{94.3} & \textbf{0.379} & \textbf{97.1} & \textbf{96.3}  \\
    \red{Ours+ResUNET} & \textbf{94.4} & \textbf{0.376} & \textbf{97.1} & \textbf{96.6}  \\
    \red{Ours+DenseUNET} & \textbf{94.4} & \textbf{0.375} & \textbf{97.2} & \textbf{96.6}  \\
    \red{Ours+H-DenseUNET} & \textbf{94.5} & \textbf{0.374} & \textbf{97.2} & \textbf{96.5}  \\
\hline
\end{tabular}
\end{center}
\caption{Performance comparison on the HNA dataset among state-of-the-art segmentation algorithms.}\label{tab:hhousesota}
\end{table}

\subsection{Comparison with the State of the Art}
We compared our proposed model with existing state-of-the-art algorithms for vessel segmentation on the three datasets. The methods in these comparisons include DDT~\cite{wang2020deep}, DVS~\cite{shin2019deep}, UNET3d~\cite{UNET3d}, nnUNET~\cite{isensee2018nnu}, ResUNET~\cite{zhang2018road}, DenseUNET~\cite{li2018h}, PSP-Net~\cite{zhao2017pyramid} and HMSA~\cite{tao2020hierarchical}. DDT performs tubular structure modeling and is specifically designed for vessel segmentation. For medical image analysis, nnUNET is considered a strong baseline as it achieves state-of-the-art performance on many well-established segmentation challenges. PSP-Net~\cite{zhao2017pyramid} and HMSA~\cite{tao2020hierarchical} are included for comparison since they are state-of-the-art methods for generic semantic segmentation. In addition, we include DVS for comparison since it also uses a GCN for structure modeling.
\red{Since the proposed framework is not limited to a specific backbone network, we integrate it with more powerful backbones, e.g. ResUNET, DenseUNET and H-DenseUNET. As shown in Table~\ref{tab:miccaisota}, \ref{tab:vhousesota} and \ref{tab:hhousesota}, the performance can be improved.}

As shown in Table~\ref{tab:miccaisota}, the proposed method achieves the state-of-the-art performance in terms of four evaluation metrics on the ASOCA dataset, and outperforms the top-6 methods in the challenge leaderboard. Specifically, our method achieves 89.91\% DICE, 0.530 ASSD 95.8\% SP and 96.0\% SR. The DICE of our method is higher than that of DDT and the top-1 method in the leaderboard by around 1.5\%.


On the ACA and HNA datasets, the proposed method also achieves the best performance among all the methods considered in the comparisons. Specifically, the proposed method outperforms DVS by 4.7\% and 3.2\% on ACA and HNA respectively in terms of DICE. This demonstrates that multi-scale feature interaction between CNNs and GCNs is important for vessel structure modeling. 

\red{The above experiments demonstrate the superiority of the proposed method on three vessel segmentation tasks. Compared to other methods~\cite{UNET3d,isensee2018nnu,zhang2018road,zhao2017pyramid,tao2020hierarchical}, the main advantage of our approach is that it constructs a vessel graph to capture the 3D structure of vessels. On the basis of the constructed vessel graph, our proposed method uses GCNs to enhance feature propagation along vessel structures, and improve the interconnection between isolated vessel predictions. Although DVS~\cite{shin2019deep} and our method both exploit GCNs, the major distinction is that we use a super-pixel algorithm to generate graph nodes from a preliminary segmentation and the pixel values of the input image. Leveraging super-pixels makes our constructed graph more completely cover potential vessel regions, and therefore, improve the skeleton recall. In addition, we make use of forward and backward feature mappings to perform more thorough feature fusion between the CNN-based UNET and the graph UNET.
}


\begin{table}[t]
\begin{center}
\begin{tabular}{ll c c c c}
\hline
Method & DICE (\%) & ASSD & SP (\%) & SR (\%)\\
\hline
\hline
    DDT \cite{wang2020deep} & 87.1 & 0.511 & 92.0 & 88.2\\
	DVS \cite{shin2019deep}  & 86.2 & 0.544 & 91.1 & 87.3  \\
    UNET3d \cite{UNET3d} & 86.0 & 0.722 & 91.0 & 87.4 \\
    \red{ResUNET} \cite{zhang2018road} & 86.4 & 0.712 & 91.2 & 87.6 \\
    \red{DenseUNET} \cite{li2018h} & 86.5 & 0.701 & 91.4 & 87.6 \\
    \red{H-DenseUNET} \cite{li2018h} & 87.1 & 0.690 & 91.6 & 87.9 \\
    nnUNET \cite{isensee2018nnu} & 87.2 & 0.631 & 92.5 & 86.6  \\
	PSP-Net \cite{zhao2017pyramid}  & 84.0 & 0.742 & 91.0 & 84.6  \\
	HMSA \cite{tao2020hierarchical} & 85.2 & 0.762 & 89.7 & 85.2  \\
    Ours & \textbf{92.1} & \textbf{0.453} & \textbf{96.4} & \textbf{93.3}  \\
    \red{Ours+ResUNET} & \textbf{92.2} & \textbf{0.451} & \textbf{96.5} & \textbf{93.5}  \\
    \red{Ours+DenseUNET} & \textbf{92.4} & \textbf{0.450} & \textbf{96.5} & \textbf{93.6}  \\
    \red{Ours+H-DenseUNET} & \textbf{92.5} & \textbf{0.448} & \textbf{96.6} & \textbf{93.6}  \\
\hline
\end{tabular}
\end{center}
\caption{Performance comparison on a subset of hard samples from the ACA dataset among state-of-the-art segmentation algorithms. Arteries in these hard samples have calcifications,  stents or tortuous segments.}\label{tab:hardvhousesota}
\begin{center}
\begin{tabular}{ll c c c c}
\hline
Method & DICE (\%) & ASSD & SP (\%) & SR (\%)\\
\hline
\hline
    DDT \cite{wang2020deep} & 88.4 & 0.504 & 94.1 & 91.3\\
	DVS \cite{shin2019deep}  & 87.3 & 0.584 & 93.2 & 90.4  \\
    UNET3d \cite{UNET3d} & 86.4 & 0.654 & 92.1 & 89.5 \\
    \red{ResUNET} \cite{zhang2018road} & 86.6 & 0.631 & 92.6 & 89.9 \\
    \red{DenseUNET} \cite{li2018h} & 86.7 & 0.598 & 92.7 & 89.9 \\
    \red{H-DenseUNET} \cite{li2018h} & 86.9 & 0.552 & 92.9 & 89.9 \\
    nnUNET \cite{isensee2018nnu} & 87.9 & 0.554 & 93.3 & 90.5  \\
	PSP-Net \cite{zhao2017pyramid}  & 87.1 & 0.567 & 92.7 & 89.0  \\
	HMSA \cite{tao2020hierarchical}  & 88.4 & 0.542 & 92.1 & 91.3  \\
    Ours & \textbf{90.1} & \textbf{0.449} & \textbf{96.0} & \textbf{94.2}  \\
    \red{Ours+ResUNet} & \textbf{90.6} & \textbf{0.441} & \textbf{96.2} & \textbf{94.3}  \\
    \red{Ours+DenseUNet} & \textbf{90.6} & \textbf{0.440} & \textbf{96.1} & \textbf{94.5}  \\
    \red{Ours+H-DenseUNet}& \textbf{90.8} & \textbf{0.432} & \textbf{96.3} & \textbf{94.6}  \\
\hline
\end{tabular}
\end{center}
\caption{Performance comparison on a subset of hard samples from the HNA dataset among state-of-the-art segmentation algorithms. Arteries in these hard samples have calcifications or tortuous segments.}\label{tab:hardhhousesota}
\end{table}

\begin{table}
\begin{center}
\begin{tabular}{ll c c c }
\hline
Method & ASOCA & ACA & HNA \\
\hline
\hline
    DDT \cite{wang2020deep} & 0.182 & 0.184 & 0.187\\
    DVS \cite{shin2019deep}  & 0.186   & 0.187  & 0.190  \\
    UNET3d \cite{UNET3d} & 0.132  & 0.134 & 0.136 \\
    nnUNET \cite{isensee2018nnu} & 0.201  & 0.204 &  0.206 \\
    \red{ResUNET} \cite{zhang2018road} & 0.136  & 0.137 & 0.139 \\
    \red{DenseUNET} \cite{li2018h} & 0.141  & 0.144 & 0.147 \\
    \red{H-DenseUNET} \cite{li2018h} & 0.139  & 0.142 & 0.146 \\
	PSP-Net \cite{zhao2017pyramid}  & 0.142  & 0.144 &  0.145 \\
	HMSA \cite{tao2020hierarchical} & 0.341  &  0.344 & 0.347  \\
    Ours & 0.190 & 0.193 &  0.198  \\
    \red{Ours+ResUNET}  & 0.192 & 0.195 &  0.201 \\
    \red{Ours+DenseUNET}  & 0.196 & 0.198 &  0.204 \\
    \red{Ours+H-DenseUNET}  & 0.195 & 0.197 &  0.202 \\
\hline
\end{tabular}
\end{center}
\caption{\red{Comparison of inference time among state-of-the-art segmentation algorithms on the ASOCA, ACA and HNA datasets. The average inference time of each algorithm on each dataset is shown. The unit is second per sample.}}
\label{tab:time}
\end{table}

To further validate the robustness of the proposed method, we collect two subsets of 35 hard samples from the test sets of ACA and HNA, respectively. Arteries in the chosen samples have calcifications, stents or tortuous segments, which significantly increase the difficulty of vessel segmentation in clinical practice. Experimental results in Table~\ref{tab:hardvhousesota} and Table~\ref{tab:hardhhousesota} show that the proposed method performs the best on these two subsets, which demonstrates the robustness of the proposed method on hard samples.

\red{Furthermore, we compare the inference time complexity of state-of-the-art networks in Table~\ref{tab:time}. As shown in the table, the inference time of our method for a computed tomography angiography image is 0.190/0.193/0.198 second on the ASOCA, ACA and HNA datasets, respectively. Since we use a GPU-based implementation~\cite{SLIC-CUDA} of the SLIC algorithm to generate super-pixels, the graph construction step of our method is very efficient, and the overall inference time of our method is comparable to that of other methods.}

\begin{table*}[t]
\begin{center}
\begin{tabular}{c|c|c|c|cccc}
\hline
UNET-0 & UNET-G & Cfeature in UNET-G & Graph Convolution $\Omega$   & DICE (\%) & ASSD & SP (\%) & SR (\%)\\
\hline
\hline
$\checkmark$ & $\checkmark$ & $\checkmark$  & $\checkmark$   & \textbf{94.2} & \textbf{0.448} & \textbf{97.1} & \textbf{95.1}  \\
$\checkmark$ & $\checkmark$ & $\checkmark$  & $\boxtimes$    & 93.1 & 0.469 & 95.3 & 94.2  \\
$\checkmark$ & $\checkmark$ & $\boxtimes$   & $\checkmark$   & 92.8 & 0.487 & 95.1 & 94.0  \\
- & $\boxtimes$  & -  & -  & 92.2 & 0.469 & 95.3 & 94.2   \\
$\boxtimes$ & $\checkmark$ & $\checkmark$  & $\checkmark$   & 92.4 & 0.470 & 95.9 & 92.7  \\
\hline
\end{tabular}
\end{center}
\caption{Effectiveness of Different Components on the ACA Dataset. `UNET-G' means the graph UNET structure on our model. If it is removed, our framework is degenerated into a cascaded model with two CNN-UNET structures. `Cfeature in UNET-G' means we fuse CNN features of different stages into UNET-G . If it is discarded, the features of UNET-G are only acquired from its first graph features $E_{1}^{g}$ that is acquired by conducting forward mapping $f$ on $E_{1}^{c}$. `Graph Convolution $\Omega$' aims to propagate message and fuse the CNN features into UNET-G. We utilize it to compare the importance of the vessel graph modelling ability.}\label{tab:aca-fuseabl}
\end{table*}

\begin{table*}[t]
\begin{center}
\begin{tabular}{c|c|c|c|cccc}
\hline
UNET-0 & UNET-G & Cfeature in UNET-G & Graph Convolution $\Omega$    & DICE (\%) & ASSD & SP (\%) & SR (\%)\\
\hline
\hline
$\checkmark$ & $\checkmark$ & $\checkmark$  & $\checkmark$   & \textbf{94.3} & \textbf{0.379} & \textbf{97.1} & \textbf{96.3}  \\
$\checkmark$ & $\checkmark$ & $\checkmark$  & $\boxtimes$  & 93.1 & 0.412 & 96.3 & 95.4 \\
$\checkmark$ & $\checkmark$ & $\boxtimes$   & $\checkmark$   & 93.0 & 0.434 & 96.1 & 95.2  \\
- & $\boxtimes$  & -  & -  & 92.4 & 0.471 & 95.9 & 94.8  \\
$\boxtimes$  & $\checkmark$ & $\checkmark$  & $\checkmark$   & 92.6 & 0.462 & 96.1 & 94.9   \\
\hline
\hline
\end{tabular}
\end{center}
\caption{Effectiveness of Different Components on the HNA Dataset.}\label{tab:hna-fuseabl}
\end{table*}

\begin{table*}[t]
\begin{center}
\begin{tabular}{c|c|c|c|cccc}
\hline
UNET-0 & UNET-G & Cfeature in UNET-G & Graph Convolution $\Omega$    & DICE (\%) & ASSD & SP (\%) & SR (\%)\\
\hline
\hline
$\checkmark$ & $\checkmark$ & $\checkmark$  & $\checkmark$   & \textbf{89.89} & \textbf{0.544} & \textbf{95.6} & \textbf{95.9}  \\
$\checkmark$ & $\checkmark$ & $\checkmark$  & $\boxtimes$  & 88.03 & 0.567 & 95.0 & 94.7 \\
$\checkmark$ & $\checkmark$ & $\boxtimes$   & $\checkmark$   & 88.01 & 0.568 & 94.8 & 94.4  \\
- & $\boxtimes$  & -  & -  & 87.12 & 0.579 & 94.8 & 93.4  \\
$\boxtimes$  & $\checkmark$ & $\checkmark$  & $\checkmark$   & 87.91 & 0.573 & 94.9 & 93.8   \\
\hline
\hline
\end{tabular}
\end{center}
\caption{Effectiveness of Different Components on the ASOCA Dataset.}\label{tab:miccai-fuseabl}
\end{table*}

\begin{table}[t]
\begin{center}
\begin{tabular}{cccccccc}
\hline
$d_{gray}$ & $d_{dis}$ & \multicolumn{2}{c}{$d_{geo}$} & DICE & ASSD & SP & SR \\
\hline
 -& -& $X^{0}$ & $A^{0}$ &-  &- &-  & -\\
\hline
\hline
$\checkmark$ & $\checkmark$  &$\checkmark$&$\checkmark$ &  \textbf{94.2} & \textbf{0.448} & \textbf{97.1} & \textbf{95.1}   \\
$\boxtimes$  & $\checkmark$ & $\checkmark$&$\checkmark$  &93.2 & 0.462 & 95.3 & 94.2  \\
$\checkmark$ & $\boxtimes$  & $\checkmark$&$\checkmark$  &93.3 & 0.465 & 95.2 & 93.1  \\
$\checkmark$ & $\checkmark$ & $\boxtimes$ &$\checkmark$ &93.0 & 0.478 & 94.1 & 92.2  \\
$\checkmark$ &  $\checkmark$   &$\checkmark$&$\boxtimes$  &93.6 & 0.487 & 95.1 & 94.2  \\
$\boxtimes$ & $\boxtimes$  & $\checkmark$ &  $\checkmark$   &93.2 & 0.488 & 94.2 & 93.3  \\
$\boxtimes$ &  $\checkmark$   &$\boxtimes$ & $\checkmark$   &93.7 & 0.481 & 95.0 & 94.3  \\
$\boxtimes$ & $\checkmark$ &  $\checkmark$   & $\boxtimes$  &93.8 & 0.472 & 95.3 & 94.0  \\
$\checkmark$ &$\boxtimes$ & $\boxtimes$ & $\checkmark$  &92.9 & 0.512 & 94.0 & 94.7  \\
$\checkmark$ &$\boxtimes$ & $\checkmark$ & $\boxtimes$  &92.8 & 0.522 & 93.7 & 94.2  \\
$\checkmark$ &  $\checkmark$   &$\boxtimes$ & $\boxtimes$  &92.8 & 0.513 & 94.1 & 94.0  \\
$\boxtimes$ &  $\boxtimes$ & $\boxtimes$ & $\checkmark$ &92.1 & 0.552 & 93.2 & 93.5  \\
$\boxtimes$ &  $\boxtimes$ & $\checkmark$   & $\boxtimes$  &92.0 & 0.561 & 93.1 & 93.2  \\
$\boxtimes$ &  $\checkmark$   &$\boxtimes$ & $\boxtimes$  &92.3 & 0.557 & 93.0 & 93.1  \\
$\checkmark$ &  $\boxtimes$ &$\boxtimes$ & $\boxtimes$  &92.1 & 0.562 & 93.4 & 93.3  \\
$\boxtimes$ & $\boxtimes$  &$\boxtimes$ & $\boxtimes$  & 92.1 & 0.541 & 95.1 & 92.1 \\
\hline
\end{tabular}
\end{center}
\caption{Effectiveness of graph node set construction on the ACA Dataset. We remove different components of graph nodes to explore their influence on our framework. Note that DICE, SP and SR are presented as percentage.}\label{tab:nodeabl}
\end{table}

\begin{table}[t]
\begin{center}
\begin{tabular}{ccccccc}
\hline
$e_{w}^{s}$ & \multicolumn{2}{c}{$e_{w}^{a}$} & DICE (\%) & ASSD & SP (\%) & SR (\%)\\
\hline
 -&  $Y^{0}$ & $A^{0}$ & - & -& - &- \\
\hline
\hline
$\checkmark$ & $\checkmark$ & $\checkmark$&   \textbf{94.2} & \textbf{0.448} & \textbf{97.1} & \textbf{95.1}   \\
$\checkmark$ & $\checkmark$ & $\boxtimes$ &93.6 & 0.465 & 91.3 & 90.7  \\
$\checkmark$ & $\boxtimes$  & $\checkmark$&93.9 & 0.462 & 91.6 & 90.1  \\
$\boxtimes$  & $\checkmark$ & $\checkmark$&93.7 & 0.461 & 91.0 & 89.8  \\
$\checkmark$ & $\boxtimes$  & $\boxtimes$ &92.4 & 0.466 & 90.2 & 88.8  \\
$\boxtimes$  & $\checkmark$  & $\boxtimes$ &92.3 & 0.472 & 90.1 & 87.4  \\
$\boxtimes$  & $\boxtimes$  & $\checkmark$ &92.2 & 0.486 & 90.3 & 88.1  \\
$\boxtimes$  & $\boxtimes$  & $\boxtimes$ &92.5 & 0.484 & 86.1 & 85.2  \\
\hline
\end{tabular}
\end{center}
\caption{Effectiveness of graph edge set construction on the ACA Dataset. We remove different components of graph edges to explore their influence on our framework. If all components are removed, graph edges become the traditional binary edges.}\label{tab:edgeabl}
\end{table}

\begin{table}[t]
\begin{center}
\begin{tabular}{r c c c c}
\hline
Method & DICE (\%) & ASSD  & SP (\%) & SR (\%)\\
\hline
\multicolumn{5}{c}{ASOCA Dataset} \\
\hline
\hline
Set1\_1 & 88.12 & 0.593 & 94.1 & 95.1  \\
Set1\_2 & 89.89 & 0.544 & 95.6 & 95.9  \\
Set1\_3 & 89.01 & 0.566 & 93.1 & 94.2  \\
Set1\_4 & 88.76 & 0.612 & 93.0 & 93.8  \\
Set2\_1 & 88.78 & 0.641 & 93.3 & 94.0  \\
Set2\_2 & 88.64 & 0.633 & 93.1 & 93.9  \\
Set2\_3 & 89.89 & 0.544 & 95.6 & 95.9  \\
Set2\_4 & 87.19 & 0.646 & 92.9 & 94.1  \\
\hline
\multicolumn{5}{c}{ACA Dataset} \\
\hline
\hline
Set1\_1 & 93.46 & 0.510 & 95.4 & 95.2  \\
Set1\_2 & 94.20 & 0.448 & 97.1 & 95.1 \\
Set1\_3 & 93.12 & 0.534 & 95.1 & 93.2  \\
Set1\_4 & 93.22 & 0.512 & 96.1 & 94.3  \\
Set2\_1 & 93.80 & 0.476 & 96.2 & 94.3  \\
Set2\_2 & 92.91 & 0.487 & 94.4 & 93.2  \\
Set2\_3 & 94.20 & 0.448 & 97.1 & 95.1  \\
Set2\_4 & 93.12 & 0.493 & 96.2 & 93.1  \\
\hline
\multicolumn{5}{c}{HNA Dataset} \\
\hline
\hline
Set1\_1 & 89.12 & 0.498 & 95.1 & 94.0  \\
Set1\_2 & 90.10 & 0.449 & 96.0 & 94.2  \\
Set1\_3 & 88.81 & 0.564 & 93.2 & 92.9  \\
Set1\_4 & 89.11 & 0.571 & 92.6 & 91.7  \\
Set2\_1 & 88.24 & 0.464 & 95.0 & 94.0  \\
Set2\_2 & 89.98 & 0.541 & 95.1 & 93.9  \\
Set2\_3 & 90.10 & 0.449 & 96.0 & 94.2  \\
Set2\_4 & 90.01 & 0.448 & 95.7 & 93.7  \\
\hline
\end{tabular}
\end{center}
\caption{\red{Performance comparison on the ASOCA, ACA, and HNA datasets among different settings of n\_segments and min\_size\_factor. Performance is measured in terms of four metrics including DICE, ASSD, SP and SR.}}\label{tab:abl_exp_super}
\end{table}

\subsection{Ablation Study}
{\flushleft \bf Ablation of graph node construction.} We investigate the effectiveness of the three components of Eqn.~(\ref{dis}) for graph node construction on the ACA dataset. As shown in Table~\ref{tab:nodeabl}, all the components play important roles in the node construction process, and $d_{geo}$ is the most important for the segmentation performance. Removing $A^{0}$ in $d_{geo}$ leads to 0.6\% performance drop and removing $X^{0}$ leads to about 1.2\% performance drop in terms of DICE.
\red{We further investigate how the hyperparameter `n\_segments' and `min\_size\_factor' of the SLIC algorithm affect the performance of our method. For the ablation study on `n\_segments', we first fix `min\_size\_factor' to 0.5 and change the value of `n\_segments' to 28000 (Set1\_1), 14000 (Set1\_2), 7000 (Set1\_3), and 3500 (Set1\_4). Then we fix `n\_segments' to 14000 and change the value of `min\_size\_factor' to 0.3 (Set2\_1), 0.4 (Set2\_2), 0.5 (Set2\_3), and 0.6 (Set2\_4). For the above eight settings, we demonstrate how the number of nodes and edges changes in Table III and Table IV. Then we conduct an ablation study on all three datasets and report the results in Table XV. From the experimental results, we can see that our model achieves the best performance by setting `n\_segments' and `min\_size\_factor' to 14000 and 0.5, respectively. The corresponding number of nodes per image is 12000, 9600, and 13000 on the ASOCA, ACA and HNA datasets, respectively.}

{\flushleft \bf Ablation of graph edge construction.} Next, we investigate the effectiveness of the two components of graph edge construction. As shown in Table \ref{tab:edgeabl}, both $e_{w}^{s}$ and $e_{w}^{a}$  are important for vessel segmentation. Furthermore, if we discard the two edge terms and use the traditional binary edge, the performance drops by 1.7\% in terms of DICE.

{\flushleft \bf Ablation of cross-network feature fusion.}
To show the effectiveness of cross-network feature fusion, we first discard UNET-G of our proposed framework and only keep the cascaded UNET. ~\rednew{As shown in Tables~\ref{tab:aca-fuseabl}, \ref{tab:hna-fuseabl} and \ref{tab:miccai-fuseabl}, the performance drops by 2.0\%, 1.9\% and 2.77\% on ACA, HNA and ASOCA respectively, which further validates the importance of vessel structure modeling. In addition, we find that the improvement of UNET-G on ASOCA dataset is much more significant than ACA and HNA. As the training set of ASOCA only contains 40 CT images, this demonstrates that CNNs cannot well exploit the characteristics of vessels when the size of training data is small. Then we evaluate the effectiveness of using multi-scale fusion and graph convolution. Experimental results show that both components are important for vessel segmentation.}

\begin{figure}[!b]
\begin{center}
\includegraphics[width=1\linewidth]{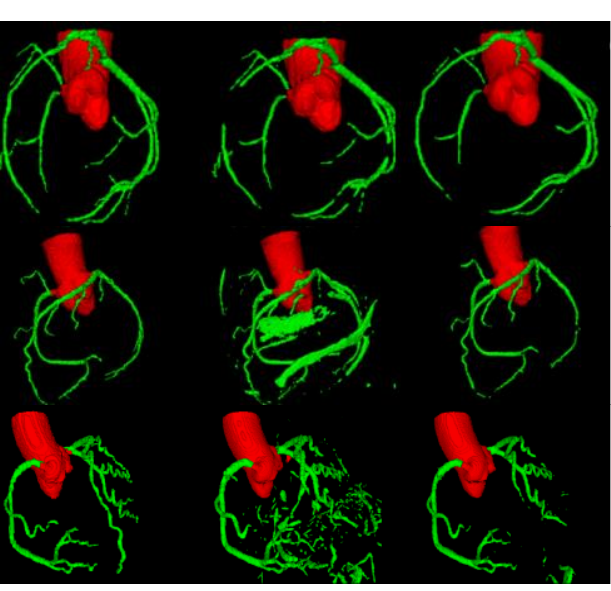}
\end{center}
\caption{Sample visual results with and without the graph module on the ACA dataset. From left to right, it is the ground truth, the result without the graph module and the result with the graph module, respectively.}\label{fig:abl-graph}
\end{figure}

\subsection{Visualization}
As shown in Fig. \ref{fig:visual}, the proposed method generates higher quality vessel masks than other state-of-the-art algorithms, including DDT, in most of the cases. Specifically, the proposed method can well exploit vessel structures and generate more complete vessel masks. In comparison to the proposed method, DDT may generate isolated segmentation masks since it is incapable of modeling the global structure of vessels. Fig.~\ref{fig:visual0} and Fig.~\ref{fig:visual3} further visualize vessel segmentation results from different methods on the ACA and HNA datasets respectively.




\red{
We add more examples for qualitative comparison in Fig.~\ref{fig:abl-graph}. The good cases show that our GCN-based cascaded network can improve vessel connectivity among individual vessel predictions and achieve a higher skeleton recall. In the meantime, most of the false positive predictions can be removed. From the bad case, we find that the proposed method is limited when the initial segmentation is far from the ground truth. In such cases, vessel segmentation errors in the initial segmentation cannot be completely corrected by our cascaded network.
}

\section{Conclusions and Future Work}


\red{In this paper, we have presented a cascaded deep neural network for vessel segmentation on CTA images. Our approach represents a new paradigm for modeling the structural information of 3D vessels using deep neural networks through the interaction between a pair of CNN-based U-Net and graph U-Net. By fusing the features across these two types of networks, our method successfully tackles the challenges brought up by the sparsity and anisotropy of vessel structures. Extensive experiments on both public and in-house datasets verify the superiority and effectiveness of our method. By constructing a vessel graph to complement CNNs, our method not only outperforms baseline methods but also achieves the state-of-the-art performance with DICE 89.91/94.8/94.5 on the ASOCA/ACA/HNA datasets, respectively.}

\red{Our proposed framework provides a stronger spatial structure representation by learning 3D vessel connectivity priors. Our future work includes 1) building a more powerful graph neural network to enhance message passing in our cross-network feature fusion module, 2) investigating better graph construction methods by exploiting more domain knowledge from medical experts, and 3) building a high-quality annotated dataset and a friendly open-source code base for 3D vessel segmentation tasks.}

\subsection*{Acknowledgment}
\textcolor{black}{
The retrospective study on our in-house datasets has been approved by the institutional review board of the Second Affiliated Hospital of Zhejiang University School of Medicine, and was carried out following the principles of the Declaration of Helsinki.
}

\bibliographystyle{IEEEtran}
\textcolor{black}{\bibliography{journalbib}}

\end{document}